\documentclass[onecolumn]{IEEEtran}
\usepackage{algorithm}
\usepackage[noend]{algpseudocode} 
\usepackage{algorithmicx}

\usepackage{amsmath,amssymb,amsfonts}
\usepackage{graphicx}
\usepackage{textcomp}
\usepackage{xcolor}
\usepackage{colortbl}
\usepackage{subfigure}
\usepackage{optidef} 
\usepackage{multicol}
\usepackage{multirow}
\usepackage{hyperref}
\usepackage{listings}
\usepackage{lipsum} 
\usepackage{cite}
\usepackage[normalem]{ulem}
\usepackage[shortlabels]{enumitem}

\makeatletter
\newcommand\ackname{Acknowledgements}
\if@titlepage
  \newenvironment{acknowledgements}{%
      \titlepage
      \null\vfil
      \@beginparpenalty\@lowpenalty
      \begin{center}%
        \bfseries \ackname
        \@endparpenalty\@M
      \end{center}}%
     {\par\vfil\null\endtitlepage}
\else
  
\fi
\makeatother

\definecolor{todo}{RGB}{190, 110, 190}

\newcommand{\done}[1]{{\color{blue}#1}}
\definecolor{sdr}{rgb}{0.0, 0.65, 0.31}

\begin{document}

\title{{\em EdgeRL}: Reinforcement Learning-driven Deep Learning Model Inference Optimization at Edge}

\author{
    \IEEEauthorblockN{Motahare Mounesan\IEEEauthorrefmark{1}, Xiaojie Zhang\IEEEauthorrefmark{2}, Saptarshi Debroy\IEEEauthorrefmark{1}}
    \\
    \IEEEauthorblockA{\IEEEauthorrefmark{1}Computer Science, City University of New York, New York, NY, USA}
    \IEEEauthorblockA{\IEEEauthorrefmark{2}Computer Science, Hunan First Normal University, Changsha, Hunan, China}
    Emails:{\{\textit{mmounesan@gradcenter.cuny.edu, xiaojie.zhang@hnfnu.edu.cn, saptarshi.debroy@hunter.cuny.edu}\}}
    }




\maketitle



\begin{abstract}

Balancing mutually diverging performance metrics, such as, processing latency, outcome accuracy, and end device energy consumption is a challenging undertaking for deep learning model  inference in ad-hoc edge environments. 
In this paper, we propose {\em EdgeRL} framework that seeks to strike such balance 
by using an Advantage Actor-Critic (A2C) Reinforcement Learning (RL) approach that can choose optimal run-time DNN inference parameters 
and aligns the performance metrics based on the application requirements. 
Using real world deep learning model and a hardware testbed, we evaluate the benefits of {\em EdgeRL} framework in terms of end device energy savings, inference accuracy improvement, and end-to-end inference latency reduction.\\
\end{abstract}

\begin{IEEEkeywords}
Edge computing, deep learning, model inference, DNN partitioning, reinforcement learning.
\end{IEEEkeywords}

\section{Introduction}
Deep learning models, particularly deep neural networks (DNN), are becoming increasingly important for mission-critical applications, such as public safety, tactical scenarios, search and rescue, and emergency triage, most of which are often edge-native. Unlike traditional edge that are typically part of the network infrastructure, a new paradigm of ad-hoc deployments of edge computing environments are currently being adopted by public safety agencies and armed forces~\cite{zhang2021effect,edgeurb,sec2023} to support mission-critical use cases. Here, heterogeneous components in the form of {\em energy-constrained} end devices (e.g., drones, robots, IoT devices) and  edge servers with varied degrees of computational and energy capacities are loosely coupled to primarily run pre-trained DNN model inference with {\em strict latency and accuracy requirements}.

In such implementations, running the entire DNN inference on end devices (e.g., drones, robots) is impractical due to their resource constraints, which makes them incapable of satisfying the inference latency and accuracy requirements. It is also not prudent to run those entire DNN models on the edge servers as they lack sufficient resource capacity (i.e., in comparison to cloud servers) that can support heterogeneous inference workloads generated from multiple end devices, simultaneously~\cite{razavi2024tale}. Thus, in recent times, an alternative approach of partial offloading/DNN partitioning/DNN splitting/collaborative inference~\cite{matsubara2022bottlefit, zhang2021effect} has gained traction that embraces segmenting the DNN models and processing the segments on end devices and edge servers collaboratively.   


However, any attempt to make such partial offloading strategy effective and practical, needs to consider the fundamental three-way trade-off between {\em end-to-end inference latency}, {\em model inference accuracy}, and {\em end device energy consumption} metrics (henceforth referred to as `latency-accuracy-energy'). This is because, each of the metrics in `latency-accuracy-energy' trade-off problem is a function of the convolutional layer of the DNN where such partition/split is carried out, as well as the unique characteristics of the involved DNNs, such as, the number of convolutional layers, the computational complexity of each layer, and the output data size at each convolutional layer of the DNN, among other things~\cite{10.1145/3093337.3037698, 10.1145/3527155, salmani2023reconciling, ghafouri2024solution}. For example, if a DNN is split at a layer whose output data size is larger, then the overall inference latency may improve due to lightweight computation at the device. However, this may result in an increase in the device energy consumption for transmitting the larger amount of output data to the edge server~\cite{zhang2021effect, zhang2023effect, icfec2024}. While, a lightweight/compressed version of a particular DNN model, when chosen to run collaboratively to lower inference latency, on the flip-side, can compromise inference accuracy due to the lower number of convolutional layers and hyperparameters in the compressed DNN.

In this paper, we address this non-trivial and grossly under-explored three-way trade-off problem by designing and developing a novel {\em EdgeRL} framework. 
The framework allows the ad-hoc edge environments select a DNN execution profile, which involves choosing an optimized version of a given DNN model from multiple pre-cached versions, whether lightweight or heavyweight, and selecting a partition cut point layer for the chosen version to perform collaborative inference with the edge server. 
This execution profile selection is framed as a Markov Decision Process (MDP) and solved using an Advantage Actor-Critic (A2C) based reinforcement learning approach. The model integrates inputs such as the end device's battery status, activity profile, available bandwidth, and kinetic activity to continuously adapt and learn system dynamics through iterative actions and rewards. The reward function aims to maximize a customizable performance metric that balances 'latency-accuracy-energy', addressing the specific requirements of different ad-hoc edge implementations. 
We validate the stability of the proposed A2C algorithm with object classification DNNs and the testbed, and rigorously examine how varying reward weights impact performance.


The rest of the paper is organized as follows. Section~\ref{sec:systemmodel} introduces the system model and solution approach. Section~\ref{evaluaton} discusses system evaluation. Section~\ref{sec:conclusion} concludes the paper.

\section{System Model and Solution Approach}
\label{sec:systemmodel}
In this section, we describe the ad-hoc edge deployment  system model, along with its energy and latency considerations, the details of the {\em EdgeRL} framework, and the RL-driven solution approach.

\subsection{System Model}
\begin{figure}[t]
    \centering
    \includegraphics[width=\linewidth]{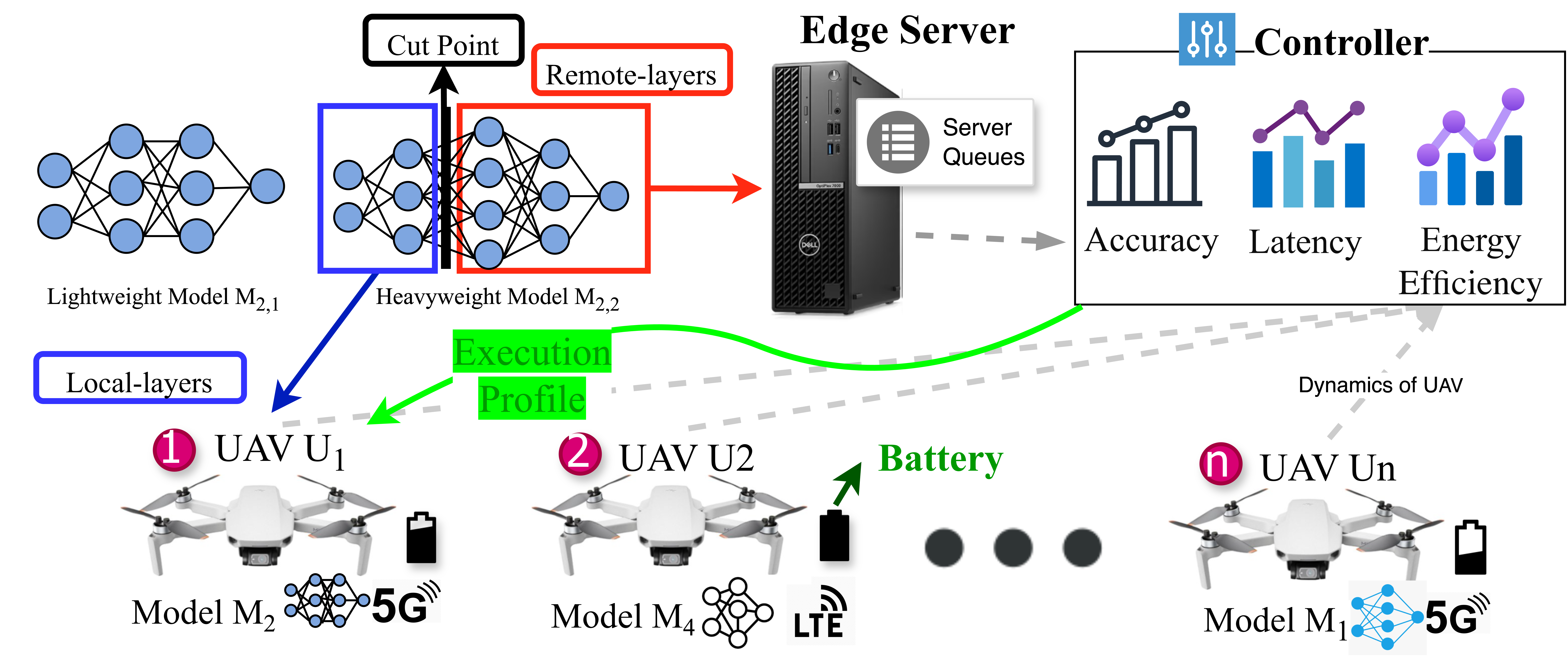}
    \caption{\footnotesize{Ad-hoc edge deployment and potential implementation of the {\em EdgeRL} framework}}
    \label{fig:systemmodel}
\end{figure}

As shown in Fig.~\ref{fig:systemmodel}, we assume an exemplary ad-hoc edge environment where end devices (UAVs in this case), in collaboration with one limited capacity edge server, perform DNN model inference for real-time missions. The components of the system are as follows:\\

\noindent \textbf{DNN Model --} We define a set of \textit{m} DNN models, denoted as $\mathcal{M} = \{M_1, M_2, ..., M_m\}$, each tailored for specific objectives and tasks towards the missions. We consider model \( M_i \) to have \( V_i \) different versions \( \{M_{i,1}, M_{i,2}, ..., M_{i,V_i}\} \), generated as a result of model optimization, with each employing either a compressed or extended architecture with diverse layers. These versions exhibit unique characteristics in accuracy and computational complexity. The accuracy and number of layers of the \textit{i}-th model in its \textit{j}-th version are expressed as $M_{i,j}^{\text{acc}}$ and $M_{i,j}^{\text{layers}}$, respectively. In this context, \( M_{i,j}^l \) denotes the `head' of the model up to and including layer \( l \) when referring to computations on the end device or local computation, and the `tail' of the model from layer \( l+1 \) onwards when referring to computations on the edge server.
Though our framework solutions are adaptable to all classes of DNNs, in this work, we focus on video processing DNNs commonly used in mission-critical applications. Consequently, we set stringent performance requirements for DNN model accuracy and latency. Specifically, each model $M_i$ must have an end-to-end inference latency not exceeding $\tau_i^{latency}$ and must achieve an accuracy of $\tau_i^{acc}$ for the application to be successful.

For the end devices, we assume realistic scenarios 
that are popular for mission critical use cases adopting ad-hoc edge environments. Specifically, we model end devices that perform other (i.e., mostly kinetic) activities on top of capturing video/image of a scene and partially performing computation. The objective is to create a realistic yet challenging device energy consumption scenario for the {\em EdgeRL} framework to address.\\

\noindent \textbf{End Device --} We define a set of \textit{n} heterogeneous Unmanned Aerial Vehicles (UAVs) or drones (very common for mission critical use cases), denoted as $\mathcal{U} = \{U_1, U_2, ..., U_n\}$, each equipped with computational capabilities. The heterogeneity comes from the UAV model that defines UAV weights and architecture, battery level, and UAV kinetic activity profile, explained later. Each UAV collaboratively executes a DNN inference task in collaboration with an edge server, facilitated through wireless connectivity between the two. Each UAV $U_k$ is defined by a quadruple \textit{(ID, build, battery level, trajectory)}, where `ID' represents the UAV's unique identification, `build' specifies the UAV model, `battery level' reflects its current power status, and `trajectory' outlines the planned path. In our model, there are three sources of device energy consumption: 1) kinetic activity, 2) computation, and 3) transmission.

Each UAV has four distinctive \textit{kinetic activities} in their flight path — forward movement, vertical ascent/descent, rotation, and hovering. Each activity generates varying energy consumption rates. For this work, we use the model in~\cite{stolaroff2018energy}.
The energy expenditure for local \textit{computation} of the head of the model (e.g. \( M_{i,j}^l \)) at UAV \( U_k \) is:
\begin{equation}
    E_{comp,i, j}^l(U_k) = P_{comp,i,j}(U_k) * T_{local,i,j}^l(U_k)
\end{equation}
\noindent where $P_{comp,i,j}$ and $T_{local,i,j}^l$ represent 
power consumption rate during the computation and the latency of executing $M_{i,j}^l$ at $U_k$, respectively.
The energy consumed in wirelessly \textit{transmitting} intermediate data generated after executing the cut point layer, along with the cut point information to the server (e.g., via WiFi or LTE), is calculated by:
\begin{equation}
    E_{trans,i,j}^l(U_k) = \beta_{k}(B) * D_{i,j}^l
\end{equation}
where $\beta_k(\text{B})$ is transmission energy consumption rate with bandwidth $B$ and $D_{i,j}^l$ is the output data size at layer \textit{l}.
For simplicity, we ignore the energy consumption for video capture. Thus, the total UAV  energy consumption is:
\begin{equation}
    E_{i,j}^l(U_k) = E_{comp,i,j}^l(U_k) + E_{trans,i,j}^l(U_k)  
\end{equation}

\noindent \textbf{Edge Server --} We consider an edge server $\mathcal{E}$ (see Fig.~\ref{fig:systemmodel}) hosted on a utility vehicle . The server has limited capacity and additional responsibilities, managing multiple end device workloads. We assume the server is continuously connected to a power source throughout the mission, so its energy consumption is excluded from our analysis.\\

\noindent{\bf Controller --} The controller, depicted in Fig.~\ref{fig:systemmodel}, is a centralized component responsible for deep learning model inference optimization and decision-making. Physically, the controller can be implemented within the edge server(s) or as a separate entity. The controller plays a pivotal role by collecting critical data such as task details, battery levels, and available transmission speeds from the UAVs, to determine optimal execution profiles, including selecting model versions and cut-point layers for each UAV. 
These execution profile decisions are then promptly communicated to the server and the UAVs, enabling them to initiate model execution based on the decision.

\subsection{Latency model}
The latency \(T_{i,j}^l\) for executing \(M_{i,j}\) collaboratively between \(U_k\) and the edge server $\mathcal{E}$, partitioned at cut point \(l\), consists of three main components. First, the \textit{local processing time} \(T_{local,i,j}^l\) represents the latency involved in processing the `head' of \(M_{i,j}^l\) at \(U_k\), constrained by the limited processing capabilities available at \(U_k\). Second, the \textit{transmission time} \(T_{trans,i,j}^l(U_k, \lambda)\) refers to the latency associated with data transfer between \(U_k\) and the edge server, determined by the transmission rate \(\lambda\). This bandwidth can be very limited depending on the mission-critical use case, often making it a limiting factor for full offloading of \(M_{i,j}^l\) to the edge server, thereby necessitating partial offloading as discussed in~\cite{zhang2021effect}. Finally, the \textit{server or remote processing time} \(T_{remote,i,j}^l\) includes both the computation time \(T_{\text{comp},i,j}^l\) for the tail of the model on the server $\mathcal{E}$ and the server queue time \(T_{queue}\). The total server processing time is given by:

\begin{equation}
    T_{remote,i,j}^l(\mathcal{E}) = T_{queue}(\mathcal{E}) + T_{comp,i,j}^l(\mathcal{E})
\end{equation}

The variability in \( T_{\text{queue}} \) at server \( \mathcal{E} \), influenced by concurrent tasks managed for other jobs by the edge server, is crucial for accurately modeling the operational dynamics of limited resource ad-hoc edge servers. Thus, 
the total end-to-end latency is:

\begin{equation}
    \scalebox{0.83}{$
    T_{i,j}^l(U_k, \lambda, \mathcal{E}) = T_{local,i,j}^l(U_k) + T_{trans,i,j}^l(U_k, \lambda) + T_{remote,i,j}^l(\mathcal{E})
    $} 
    \label{totallatency}
\end{equation}

\subsection{Deep Reinforcement Learning (DRL) Agent}
The controller trains a DRL agent to handle the dynamic nature of the system in ad-hoc edge environments. Specifically, we model the DNN optimization problem as a Markov Decision Process (MDP) and use a time-slot-based decision-making approach based on the Advantage Actor-Critic (A2C) algorithm~\cite{konda1999actor,mnih2016asynchronous}.
The choice of A2C is driven by its efficiency and effectiveness. In A2C, an agent serves both as the actor and the critic, combining policy-based and gradient-based methods. The actor makes decisions, while the critic evaluates these decisions and provides feedback to refine strategies. This collaborative approach accelerates training and enhances learning with each experience. Moreover, A2C is a stable algorithm capable of handling large observation spaces, such as environments with potentially multiple UAVs and corresponding DNN models and versions. The A2C agent operates in an environment characterized by a finite set of states denoted as $\mathcal{S}$ and a finite set of actions denoted as $\mathcal{A}$, under a time-slot based system, with intervals of $\delta$ time units. 

\textbf{$\mathcal{S}$} represents the \textbf{state space} of the environment. At time \(t\), the state \(s(t) \in S\) includes the battery level of the \(k\)th UAV (\(b_k(t)\)), task availability (\(\alpha_k(t)\)), available transmission power (\(P_k^t\)), the \textit{DNN} model (\(m_k\)), and the percentages of forward flight (\(F_k(t)\)), vertical movement (\(V_k(t)\)), and rotational movement (\(R_k(t)\)). The activity profile describes the distribution of these movements over the next \(\delta\) seconds.

\begin{align}
    \mathcal{S} =& \Big \{ 
    s(t) = [
    s_1(t), s_2(t), \ldots , s_n(t)] ~:~ \forall{k \in |\mathcal{U}|}, \nonumber \\
    &s_k(t) = (b_k(t), \alpha_k(t), P_k^t(t), m_k(t), F_k(t), V_k(t),
    R_k(t)), \nonumber \\
    & b_k(t) \in [1,10], \alpha_k(t) \in \{0,1\}
    \Big \} && \label{eq:state}
\end{align}

\noindent \textbf{$\mathcal{A}$} represents the \textbf{action space} and is a Multi-discrete space encompassing decisions for each UAV device. At time $t$, the action $a(t)$ performed by the agent determines the execution profile, i.e., the DNN version($j$) and the cut point($l$)  for model $M_{m_{k}(t)}$:

\begin{align}
    \centering
    \mathcal{A} =& \Big \{  
    a (t) = [a_1(t), \ldots, a_n(t)] ~:~ \forall{k \in |\mathcal{U}|}, \nonumber \\
    &a_k(t) = (j, l), j \in V_{m_k(t)}\thinspace\text{and}\thinspace l \in L_{m_k(t),j}
    \Big \} && \label{eq:actions}
\end{align}

\noindent The \textbf{reward function} $R(t)$ denotes the immediate reward acquired following the transition from state $s(t)$ to state $s(t + 1)$ by executing action $a(t)$. This reward is computed as a weighted average of three key system performance requirement metrics: end device energy expenditure, model accuracy, and end-to-end inference latency. Such an approach makes the solution flexible where different combinations of the weights can be designated based on system objectives and the relative priorities of performance metrics. To this end, we define separate normalized performance scores for accuracy ($\mathcal{A}$), latency ($\mathcal{L}$), and energy consumption ($\mathcal{E}$). 
Given that UAV \( U_k \) is running DNN model \( M_i \), the average reward function for an action $a_k = (j, l)$ over all the end devices can be defined as:

\begin{align}
    \centering
    \scalebox{1}{$
    R(t) = \frac{1}{|\mathcal{U}|}\sum_{k\in\mathcal{U}} 
        w_1\mathcal{A}(M_{i,j}) + w_2\mathcal{L}(M_{i,j}^l, U_k)
        + w_3\mathcal{E}(M_{i,j}^l, U_k) 
    $} && \label{eq:reward}
\end{align}


\noindent where $ \sum_{i=1}^3 w_i = 1$ and the normalized performance scores are defined as follows: 
\begin{equation}
    \mathcal{A}(M_{i,j}) = \frac{1}{1+e^{-p.(M_{i,j}^{acc}-q)}}    
\end{equation}
\begin{equation}
    \mathcal{L}(M_{i,j}^l, U_k) = 1 - \frac{T_{i,j}^l (U_k, \lambda, \mathcal{E})}{T_{local,i,j}^{L_i} (U_k)}
\end{equation}
\begin{equation}
    \mathcal{E}(M_{i,j}^l, U_k) = 
    1 - \frac{E_{i,j}^l (U_k)}{E_{i,j}^{L_i} (U_k)}
\end{equation}
A2C ~\cite{mnih2016asynchronous} model is a hybrid method comprises two neural networks: i) The Actor network which is a Policy Gradient algorithm that learns a policy $\pi$ deciding on what action to take, and ii) the Critic network which is a Q-learning algorithm offering feedback for policy enhancement. We design an online learning algorithm which runs on the centralized controller working as the system manager. During initialization phase, the agent creates actor and critic networks with randomly assigned weights. The agent then continuously interacts with the environment and makes execution profile decisions in each time slot ($\delta$). At the end of each episode, both the actor and critic networks' weights undergo weight updates with a batch of experienced transitions. Our Critic network features two fully connected layers with feature sizes of 512 and 256, respectively. To adapt the Multi-Discrete action structure, which determines the version and cut-point for each device, the actor network incorporates an additional shared layer for related action values. Specifically, every two values that correspond to each UAV device share an extra layer with a feature size of 128.

\subsection{EdgeRL Framework Design}
The proposed {\em EdgeRL} framework has a centralized controller, serving as the system manager for decision-making processes. The end devices, i.e., UAVs in our case play a crucial role by transmitting essential information such as task details, battery levels, and available transmission speeds to the controller. This aggregated data forms the system's state, which is then processed by an actor network within the controller. The actor network utilizes this information to generate actions, taking into account various factors like system performance and resource availability.
Once actions are generated, they are relayed back to the respective UAV devices. Simultaneously, the system records rewards based on performance metrics such as accuracy, latency, and energy consumption within the edge environment. Following this, a critic network estimates the advantage values and trains both the actor and critic networks based on the actions taken and the resulting rewards.
Continuing through this iterative learning process, the system refines and adapts until it reaches convergence, ensuring optimal performance and responsiveness to environmental variables. Each episode concludes when all UAV devices' batteries are depleted.

\section{Evaluation}
\label{evaluaton}
Next, we evaluate the performance of our proposed framework through hardware testbed experimental evaluation. 

\subsection{Testbed Setup and Experiment Design}
For our ad-hoc edge deployment testbed, we utilize three NVIDIA Jetson TX2 devices as computational units of the end devices/UAVs. Additionally, a Dell PowerEdge desktop with 16 cores 3.2 GHz CPU serves as the edge server. 
The network connectivity between the TX2 devices and the edge server is established through an Ettus USRP B210 acting as the access point, and can operate on both WiFi and LTE bands. Due to the lack of UAV hardware availability, we simulate UAV kinetic activity based on an average size drone \textit{UAV Systems Aurelia
X4 Standard} and compute energy consumption of each movement based on the model proposed in~\cite{stolaroff2018energy}. To account for device heterogeneity, we consider three distinct \textit{activity profiles} for UAVs, each representing varying levels of kinetic activity. For our experiment, we specifically use the High activity profile, which features a dominant forward flight rate of 80\%, with minimal vertical and rotational movements (10\% each). This profile represents the most challenging scenario, as it emphasizes extensive forward motion, which generally requires greater coverage.

For the experiments, we mostly focus on object classification tasks as exemplar video processing applications. The UAV devices, execute three popular classification DNNs, viz., VGG, ResNet, and DenseNet.
As for different versions, 
we assume that each DNN has two variants: a lightweight, less accurate model (e.g., VGG11, ResNet18, and DenseNet121), and a heavyweight, more accurate model (e.g., VGG19, ResNet50, and DenseNet161). Furthermore, drawing from insights our analysis, we identify four potential cut points for each such DNN version (Table~\ref{tab:cutpoints}), to enable collaborative DNN inference. 
Apart from the object classification jobs arriving from the UAVs, we simulate the edge server to support other mission related jobs with exponential arrival rate which impact the size of the queue to follow a Poisson point process. We use a time slot duration of $\delta = 30s$ to meet reconnaissance demands in ad-hoc edge environments, with the controller making decisions at each interval.


\begin{table}[t]
\scriptsize
\caption{Candidate cut points for the each model}
\vspace{-0.1in}
    \centering
    \begin{tabular}{lcl}
        \hline
        \textbf{Model} & \textbf{Version} & \textbf{Candidate Cut Points} \\
        \hline
        \multirow{2}{*}{VGG} & 11 & 3, 6, 11, 27 \\
        & 19 & 5, 10, 19, 43 \\
        \hline
        \multirow{2}{*}{ResNet} & 18 & 4, 15, 20, 49 \\
        & 50 & 4, 13, 20, 115 \\
        \hline
        \multirow{2}{*}{DenseNet} & 121 & 4, 6, 8, 14 \\
        & 161 & 4, 6, 8, 14 \\
        \hline
    \end{tabular}
    \label{tab:cutpoints}
\end{table}

\subsection{Reward Sensitivity Analysis}
We explore the sensitivity of the reward function for each performance metric across different DNN models. 

\subsubsection{Sensitivity of accuracy weight:}
\begin{figure}[t]
    \centering
    \subfigure[]{%
        \includegraphics[width=2.75cm]{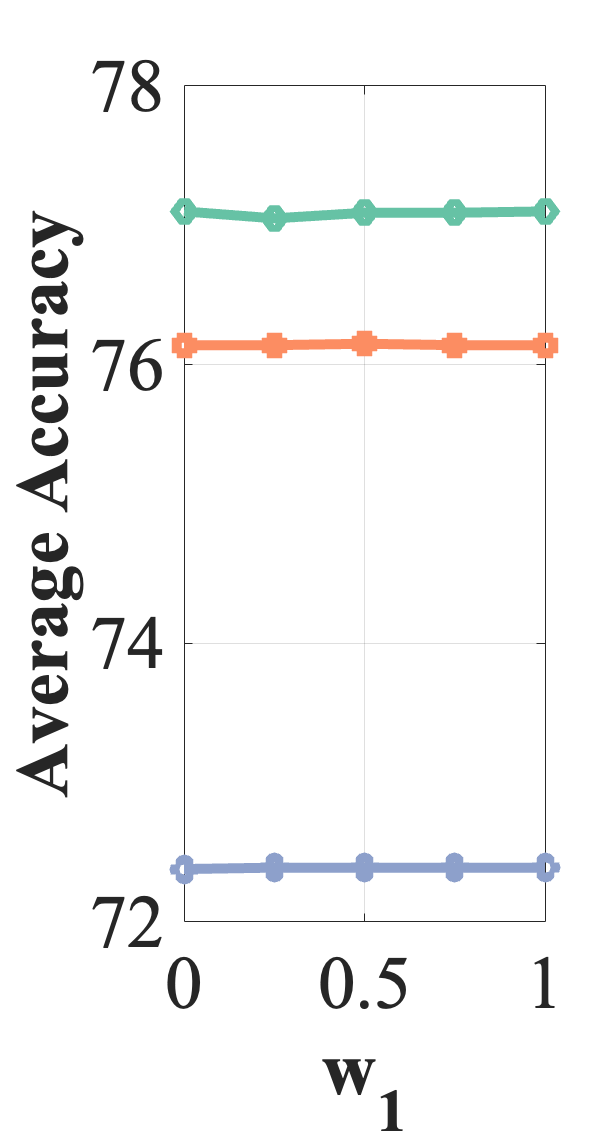}
        \label{fig:accuracy_accuracy}
    }\hspace{-0.1in}
    \subfigure[]{%
        \includegraphics[width=2.75cm]{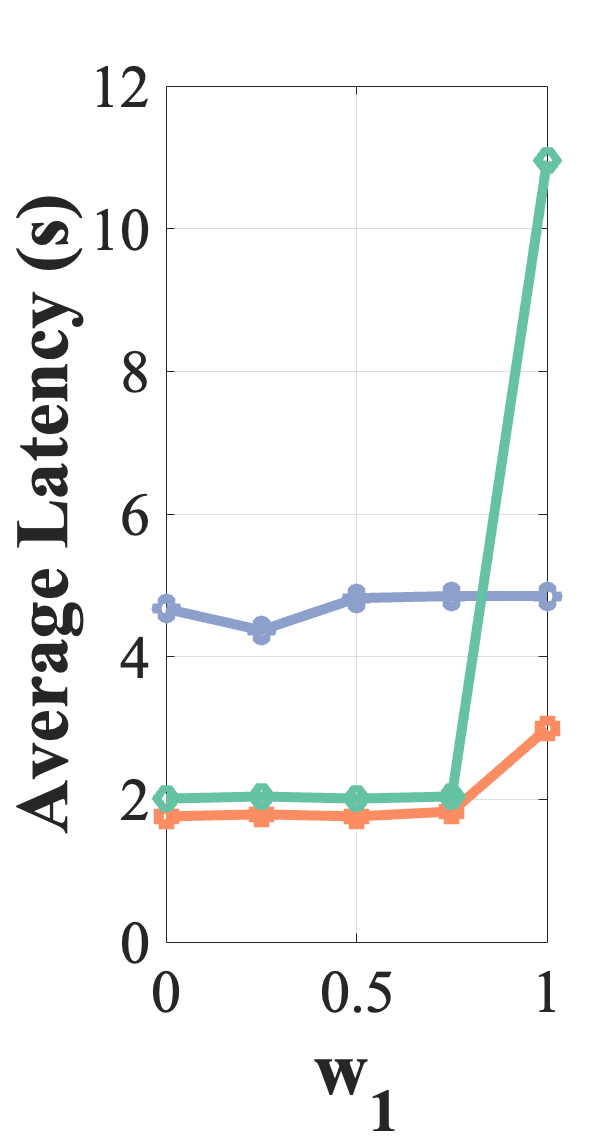}
        \label{fig:latency_accuracy}
    }\hspace{-0.1in}
    \subfigure[]{%
        \includegraphics[width=2.75cm]{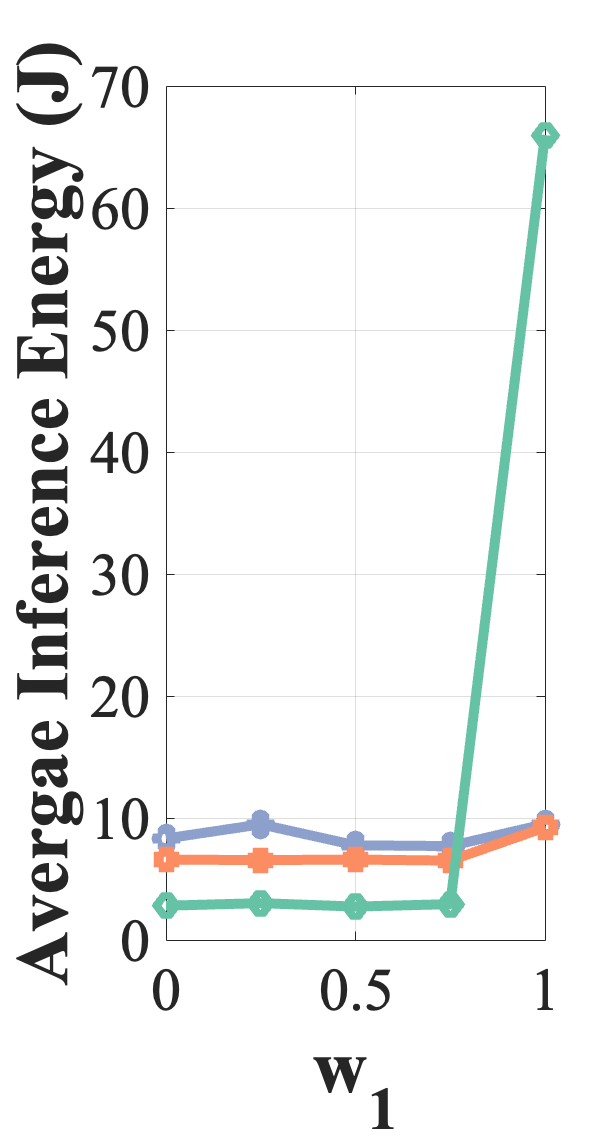}
        \label{fig:inference_accuracy}
    }\hspace{-0.1in}
    \subfigure[]{%
        \includegraphics[width=2.75cm]{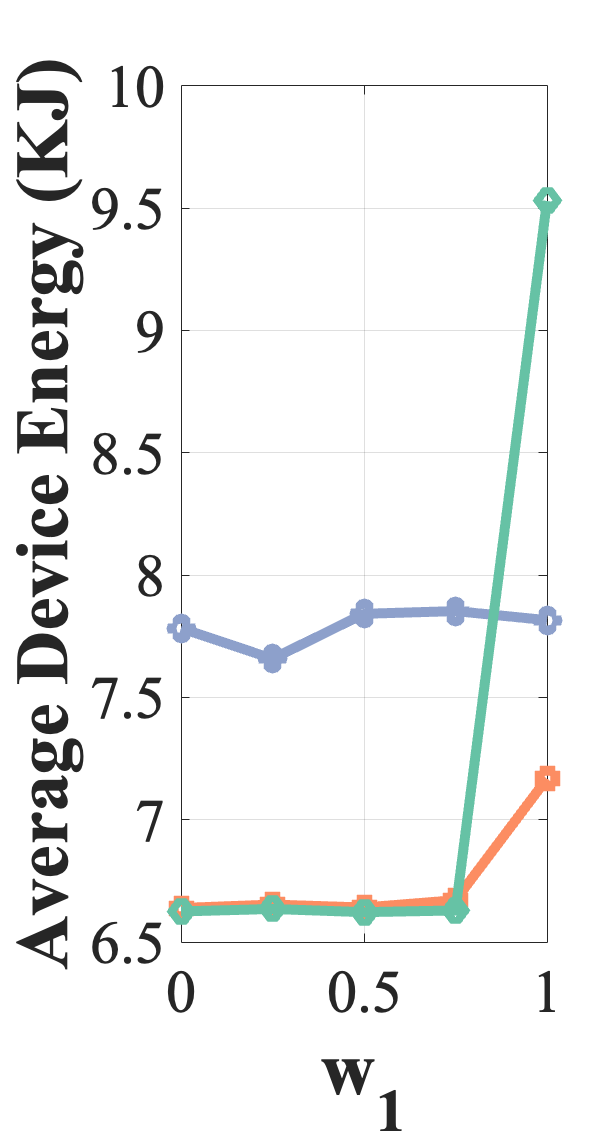}
        \label{fig:energy_accuracy}
    }\hspace{-0.1in}
    \subfigure[]{%
        \includegraphics[width=2.8cm]{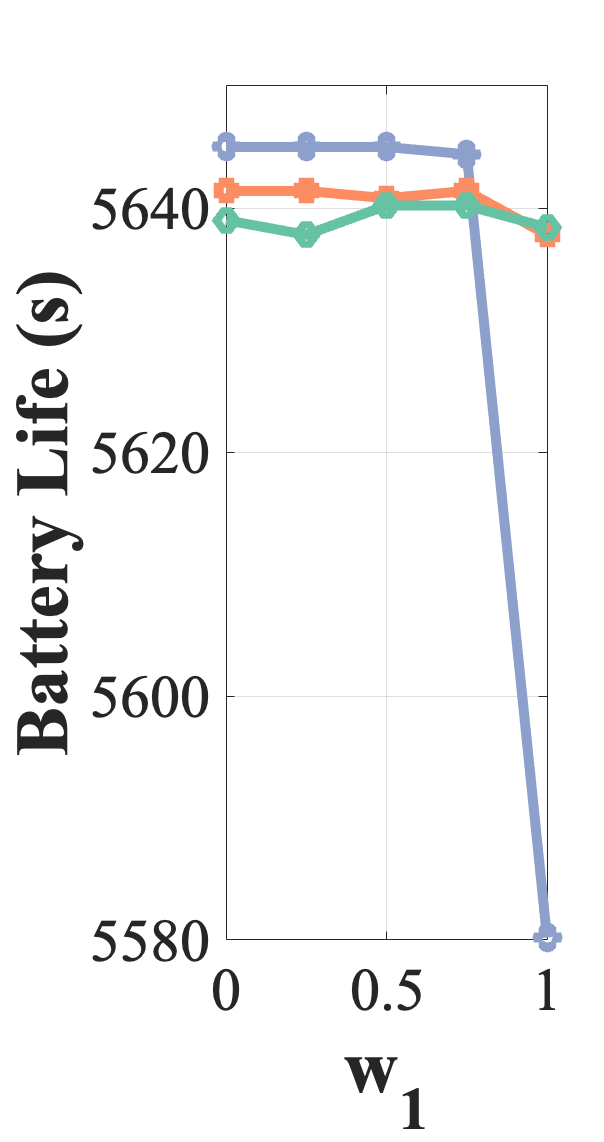}
        \label{fig:batterylife_accuracy}
    }\\
    \subfigure{%
        \includegraphics[width=5cm]{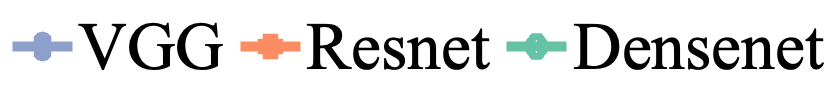}
    }\\
    \caption{System performance over varying accuracy weight}
    \label{fig:accuracyreward}
\end{figure}

Fig.~\ref{fig:accuracyreward} showcases the system performance for varying weight of accuracy reward in Eqn.~\eqref{eq:reward}. 
A notable observation (as seen in Fig.~\ref{fig:accuracy_accuracy}) is that the higher accuracy versions demonstrate better latency and energy efficiency. This is evident by the sustained high accuracy even when the accuracy reward weight is set to zero. There is minimal improvement in accuracy with the increase in accuracy weight in the reward function. Moreover, as we increase the weight, there is a noticeable decline in latency and energy performance. This trend can be attributed to the selection of cut points.

\subsubsection{Sensitivity of latency weight:}
\begin{figure}[t]
    \centering
    \subfigure[]{%
        \includegraphics[width=2.75cm]{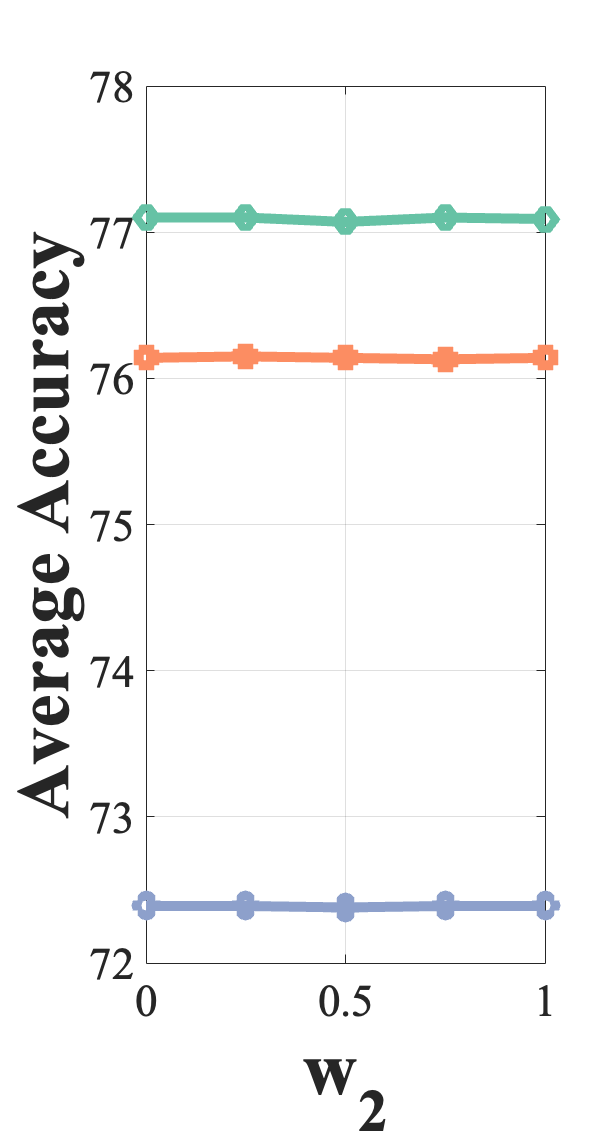}
        \label{fig:accuracy_latency}
    }\hspace{-0.1in}
    \subfigure[]{%
        \includegraphics[width=2.75cm]{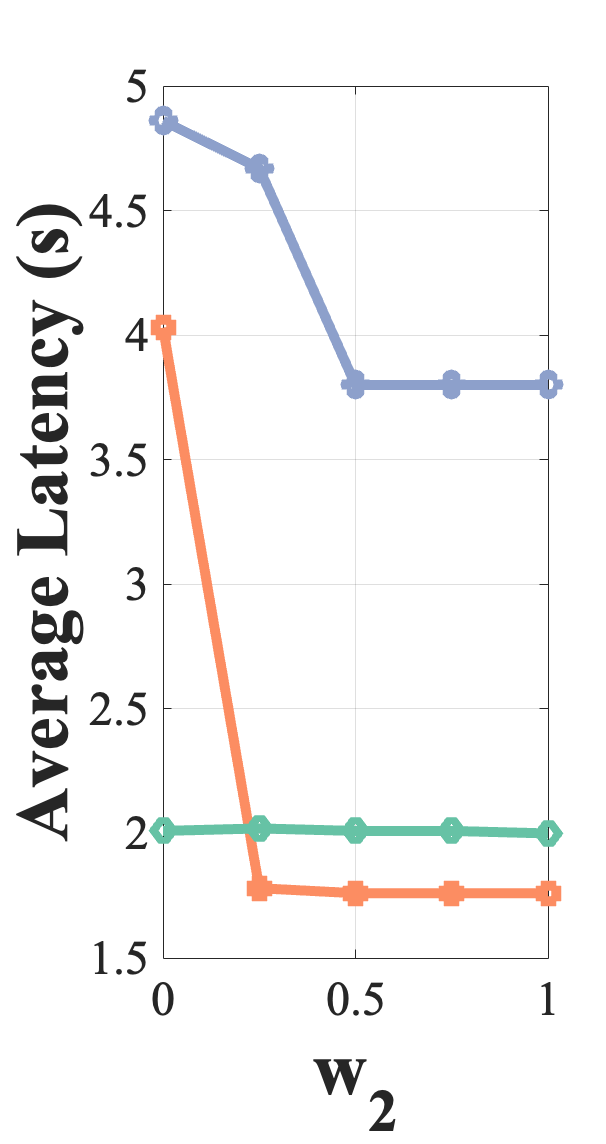}
        \label{fig:latency_latency}
    }\hspace{-0.1in}
    \subfigure[]{%
        \includegraphics[width=2.75cm]{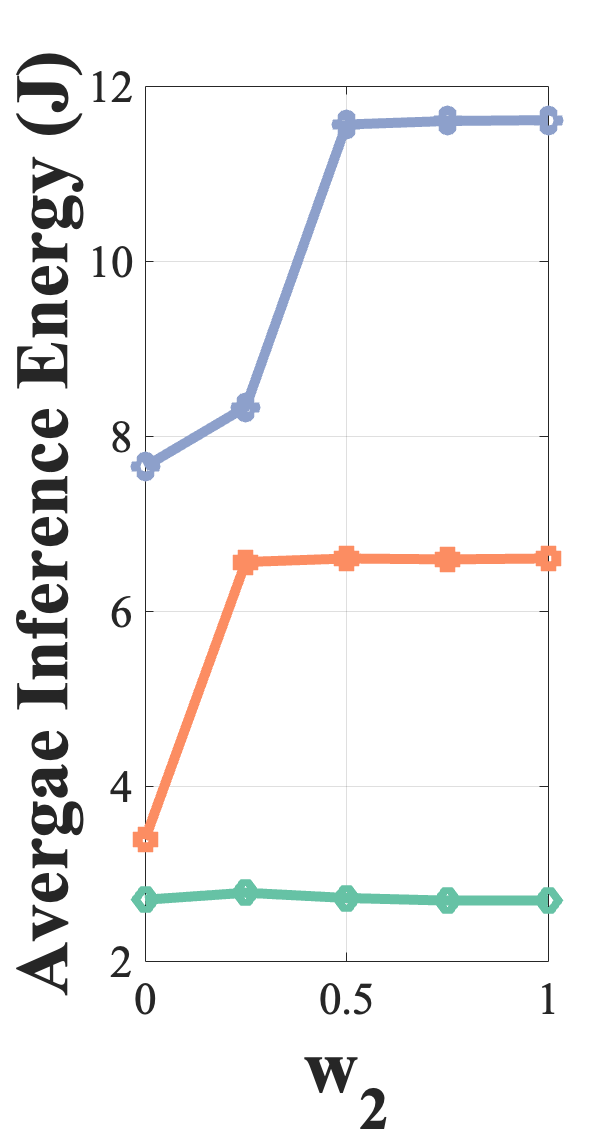}
        \label{fig:inference_latency}
    }\hspace{-0.1in}
    \subfigure[]{%
        \includegraphics[width=2.75cm]{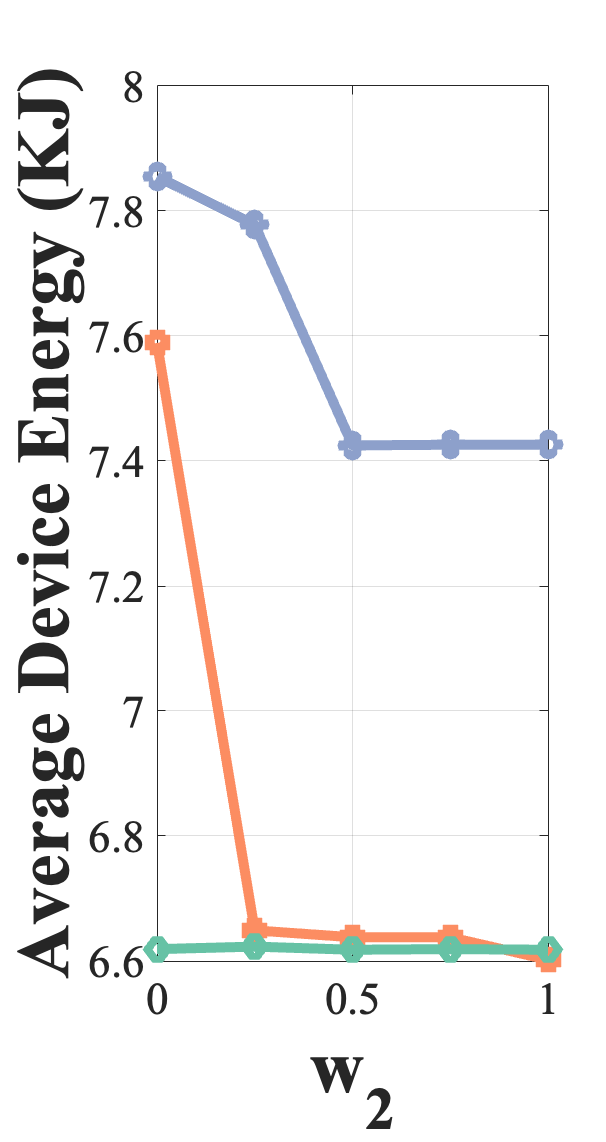}
        \label{fig:energy_latency}
    }\hspace{-0.1in}
    \subfigure[]{%
        \includegraphics[width=2.8cm]{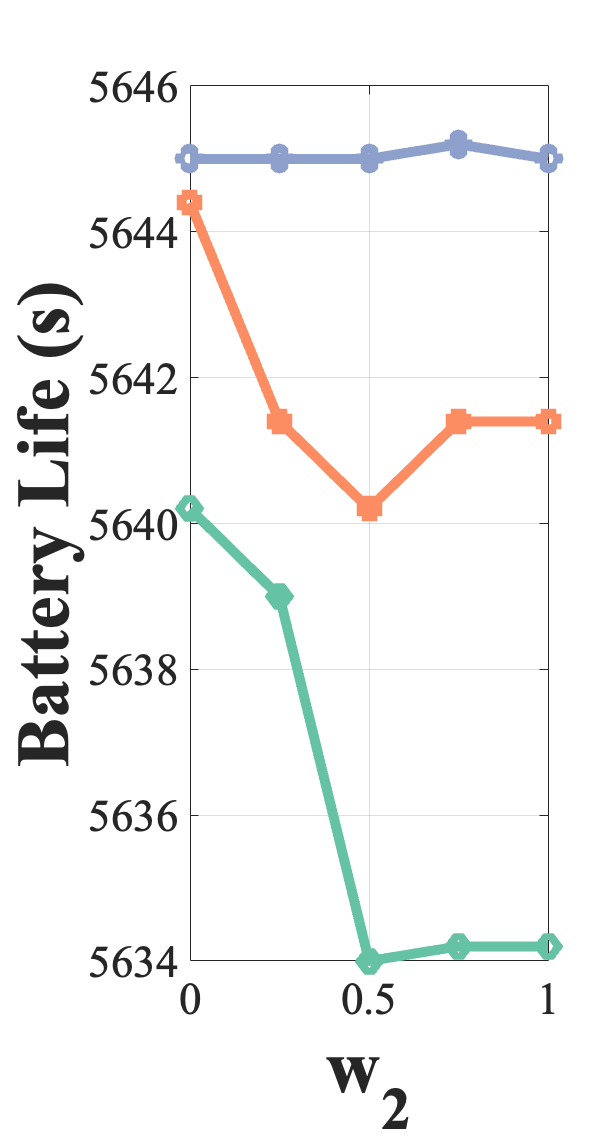}
        \label{fig:batterylife_latency}
    }\\
    \vspace{-0.15in}
    \subfigure{%
        \includegraphics[width=5cm]{Figures/EvaluationResults/ScreenShot03}
    }
    \caption{System performance over varying latency weight}
    \vspace{-0.1in}
    \label{fig:latencyreward}
\end{figure}

Fig.~\ref{fig:latencyreward} presents the findings from similar experiments with latency reward weight manipulation. 
As we increase the emphasis on latency, there's a noticeable decrease in the average latency of the models. However, this reduction comes at the cost of increased inference energy consumption. Consequently, such a pattern inevitably leads to a diminished battery life for the devices. This observation is further corroborated when we compare Figs.~\ref{fig:latency_latency} and ~\ref{fig:latency_energy} where a clear tradeoff between latency and energy consumption emerges.
The selection of versions and cut points for the $w_2=0$ and $w_2=1$ models is detailed in Tab.~\ref{tab:versioncutpoint}. Notably, with a latency weight of $0$, the energy score predominance results in a greater proportion of layers being processed remotely. However, due to latency in transmission for $w_2=1$, offloading is postponed until later layers.

\begin{table}[b]
\scriptsize
    \centering
    \caption{Cut point selection for reward weight manipulation}
    \vspace{-0.1in}
    \begin{tabular}{cccccc}
        \hline
        
        \multirow{2}{*}{Model} & \multirow{2}{*}{Version} & \multicolumn{4}{c}{Cut Point for} \\
        & & $w_2$ : 0&$w_2$ : 1 &$w_3$ : 0&$w_3$:1  \\
        \hline
        VGG & 19 & 5 & 10 & 10 & 5 \\
        ResNet & 50 & 3 & 13 & 13 & 3 \\
        DenseNet & 161 & 4 & 4 & 4 & 4 \\
        \hline
    \end{tabular}
    \label{tab:versioncutpoint}
    \vspace{-0.2in}
\end{table}

\subsubsection{Sensitivity of energy weight:}
\begin{figure}[t]
    \centering
    \subfigure[]{%
        \includegraphics[width=2.75cm]{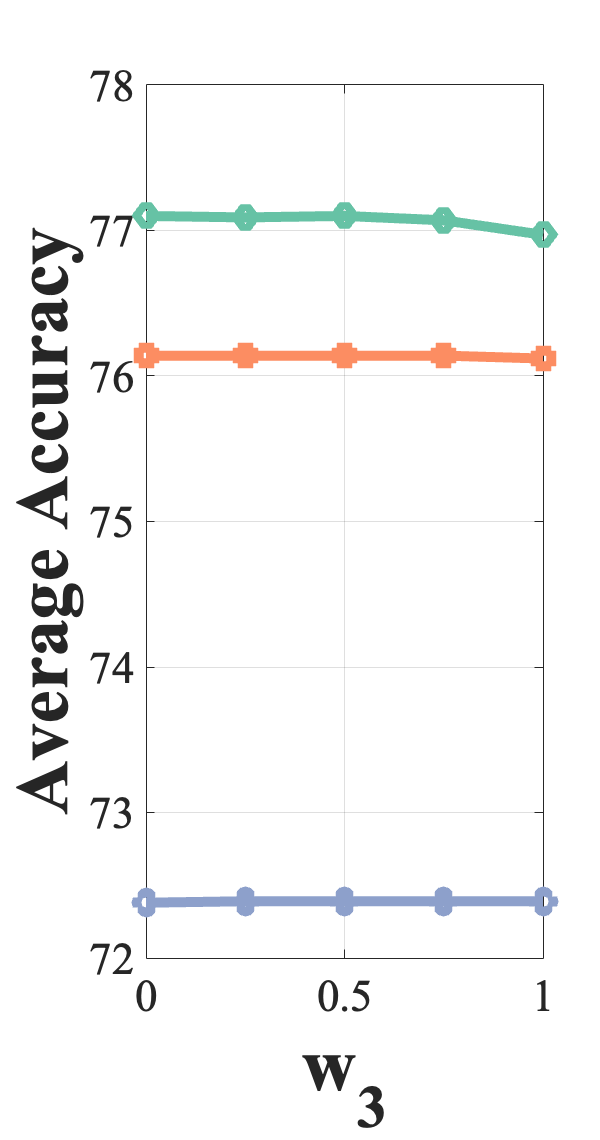}
        \label{fig:accuracy_energy}
    }\hspace{-0.1in}
    \subfigure[]{%
        \includegraphics[width=2.75cm]{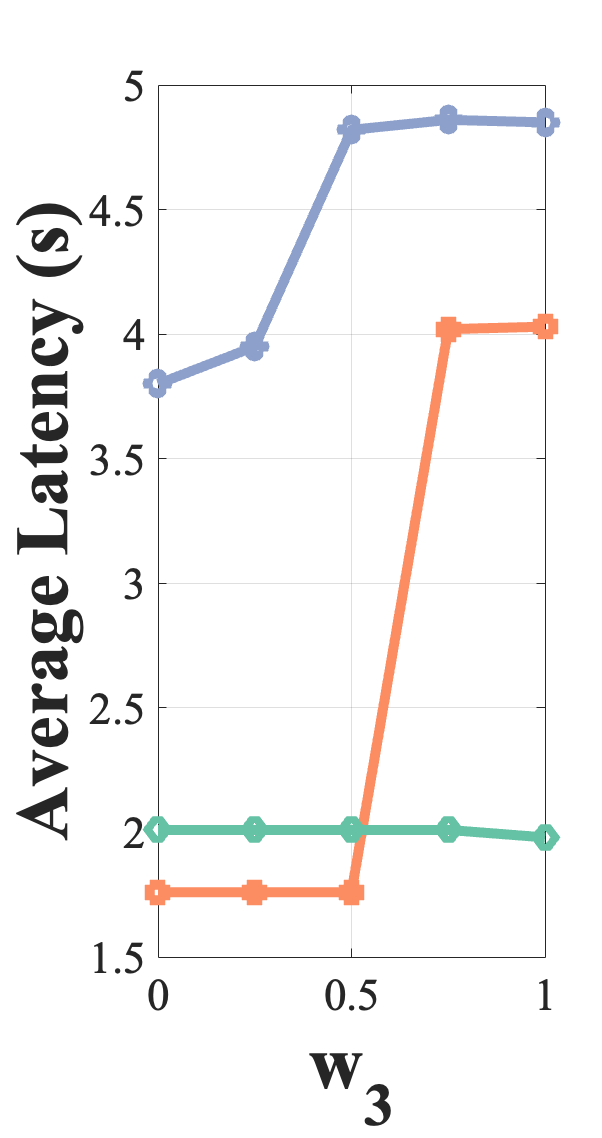}
        \label{fig:latency_energy}
    }\hspace{-0.1in}
    \subfigure[]{%
        \includegraphics[width=2.75cm]{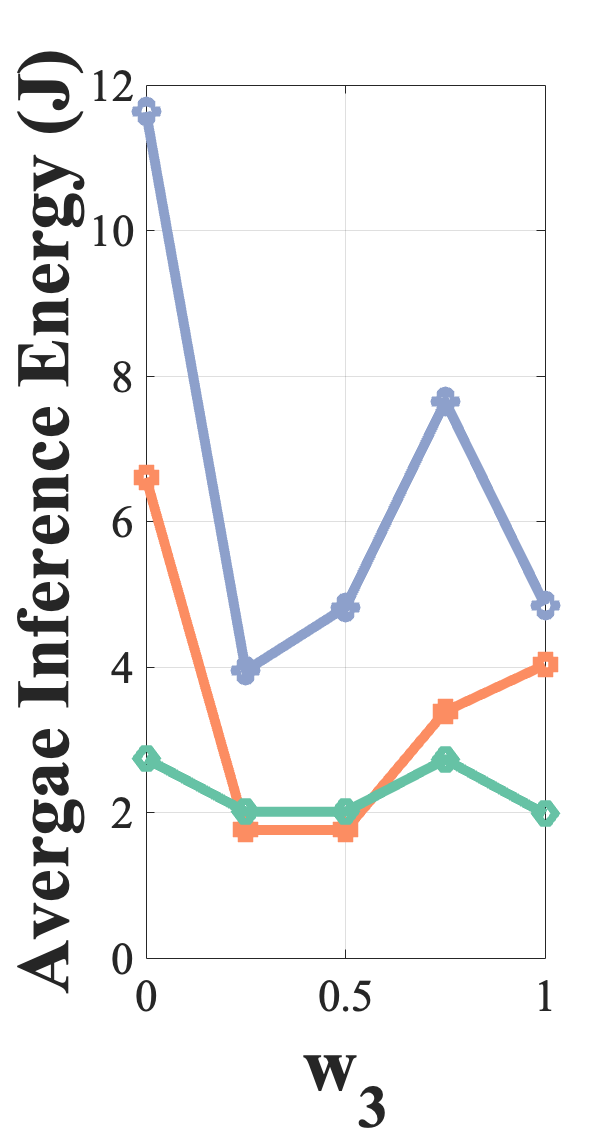}
        \label{fig:inference_energy}
    }\hspace{-0.1in}
    \subfigure[]{%
        \includegraphics[width=2.75cm]{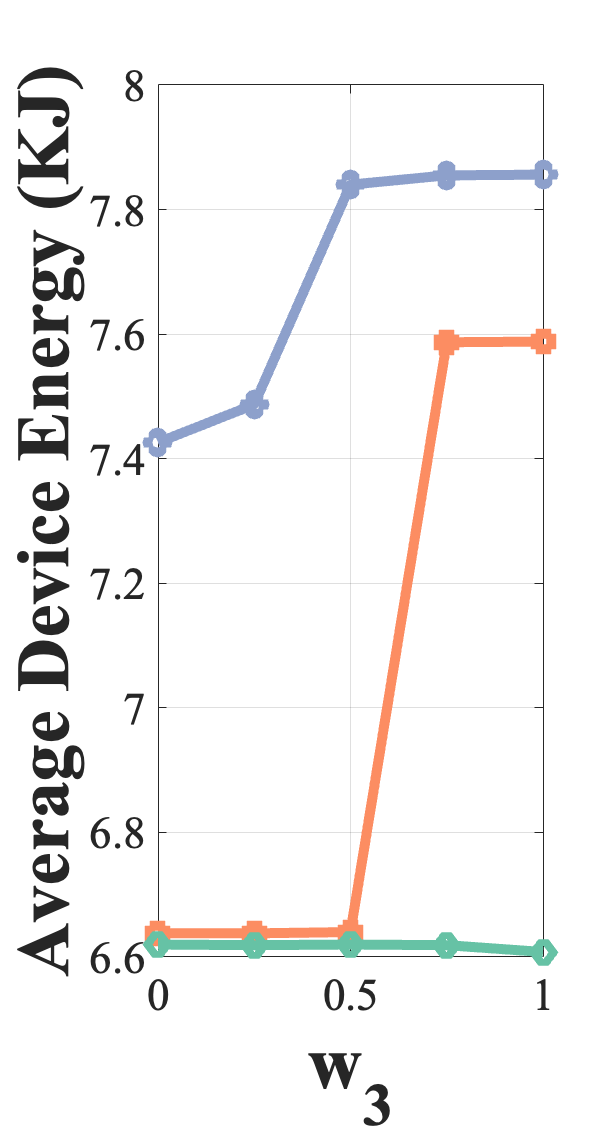}
        \label{fig:energy_energy}
    }\hspace{-0.1in}
    \subfigure[]{%
        \includegraphics[width=2.8cm]{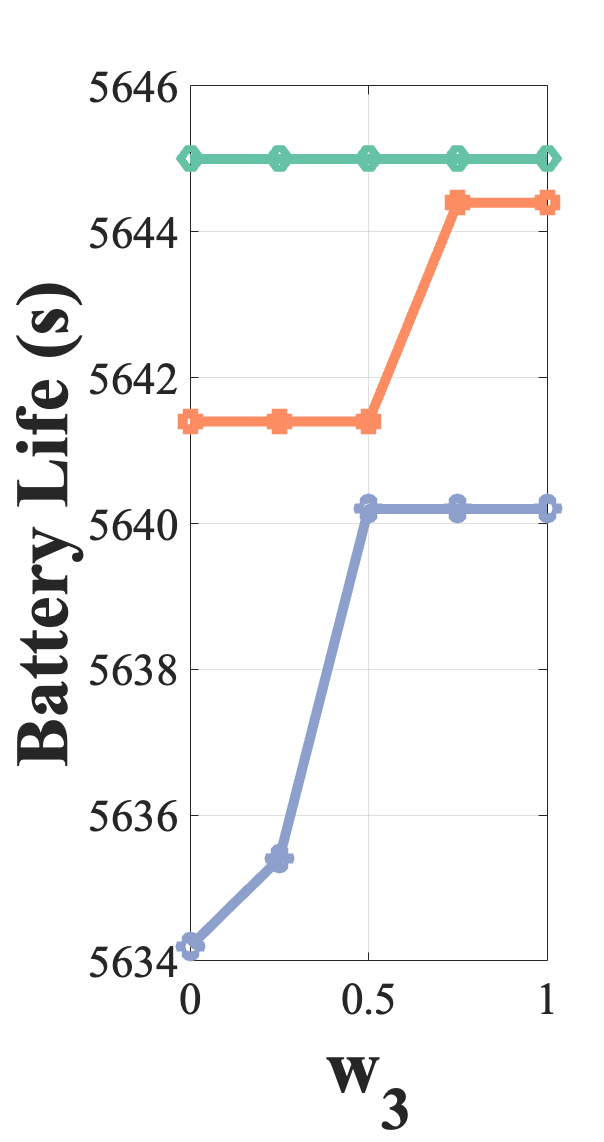}
        \label{fig:batterylife_energy}
    }\\
    \vspace{-0.15in}
    \subfigure{%
        \includegraphics[width=5cm]{Figures/EvaluationResults/ScreenShot03}
    }
    \caption{System performance over varying energy weight}
    \vspace{-0.1in}
    \label{fig:energyreward}
\end{figure}

Next, Fig.~\ref{fig:energyreward} illustrates the findings from similar experiments with energy consumption reward weight manipulation. For obvious reasons with increasing the weight, the inference energy consumption drops as energy is given the higher priority. However, end device/UAV energy consumption (from Fig.~\ref{fig:batterylife_energy}) shows a different trend, specifically, the device running \textit{DenseNet} - the most energy-consuming among the models. For accuracy performance, it can be observed that only \textit{ResNet} experiences a drop, as the energy savings in other models come from the choice of the optimal cut-point rather than the architecture (as seen in Tab.~\ref{tab:versioncutpoint}). 
It is also observed that when the energy reward is higher, the battery drains slower. E.g., the UAV running \textit{VGG} stays alive 6 more seconds when the model is energy efficient.
The choice of versions and cut points for these two models is also shown in Tab.~\ref{tab:versioncutpoint}. The interplay of energy and latency in cut point selection is evident.

\section*{Acknowledgements}
 This  work supported by the National Science Foundation under Award Number: CNS-1943338 and OAC-2232889.

\section{Conclusions}
\label{sec:conclusion}
In this paper, we analyzed the end-to-end latency vs. inference accuracy vs. device energy consumption trade-off for ad-hoc edge deployments  
and proposed {\em EdgeRL} framework that employs a novel A2C based RL model. 
The {\em EdgeRL} framework performs DNN version selection and cut point selection based on resource availability and system performance requirements. 
We demonstrated how the underlying A2C based RL agent learnt about the environment through actions and rewards which eventually converged at an optimal trade-off point for involved performance metrics. 
Using real world DNNs and a hardware testbed, we evaluated the benefits of {\em EdgeRL} in terms of device energy saving, accuracy improvement, and end-to-end latency reduction.

\bibliographystyle{IEEEtran}
\bibliography{reference}

\begin{thebibliography}{10}
\providecommand{\url}[1]{#1}
\csname url@samestyle\endcsname
\providecommand{\newblock}{\relax}
\providecommand{\bibinfo}[2]{#2}
\providecommand{\BIBentrySTDinterwordspacing}{\spaceskip=0pt\relax}
\providecommand{\BIBentryALTinterwordstretchfactor}{4}
\providecommand{\BIBentryALTinterwordspacing}{\spaceskip=\fontdimen2\font plus
\BIBentryALTinterwordstretchfactor\fontdimen3\font minus \fontdimen4\font\relax}
\providecommand{\BIBforeignlanguage}[2]{{%
\expandafter\ifx\csname l@#1\endcsname\relax
\typeout{** WARNING: IEEEtran.bst: No hyphenation pattern has been}%
\typeout{** loaded for the language `#1'. Using the pattern for}%
\typeout{** the default language instead.}%
\else
\language=\csname l@#1\endcsname
\fi
#2}}
\providecommand{\BIBdecl}{\relax}
\BIBdecl

\bibitem{zhang2021effect}
X.~Zhang, A.~Pal, and S.~Debroy, ``Effect: Energy-efficient fog computing framework for real-time video processing,'' in \emph{2021 IEEE/ACM 21st International Symposium on Cluster, Cloud and Internet Computing (CCGrid)}.\hskip 1em plus 0.5em minus 0.4em\relax IEEE, 2021, pp. 493--503.

\bibitem{edgeurb}
------, ``Edgeurb: Edge-driven unified resource broker for real-time video analytics,'' in \emph{NOMS 2024-2024 IEEE Network Operations and Management Symposium}, 2024, pp. 1--8.

\bibitem{sec2023}
X.~Zhang, H.~Gan, A.~Pal, S.~Dey, and S.~Debroy, ``On balancing latency and quality of edge-native multi-view 3d reconstruction,'' in \emph{2023 IEEE/ACM Symposium on Edge Computing (SEC)}, 2023, pp. 1--13.

\bibitem{razavi2024tale}
K.~Razavi, M.~Salmani, M.~M{\"u}hlh{\"a}user, B.~Koldehofe, and L.~Wang, ``A tale of two scales: Reconciling horizontal and vertical scaling for inference serving systems,'' \emph{arXiv preprint arXiv:2407.14843}, 2024.

\bibitem{matsubara2022bottlefit}
Y.~Matsubara, D.~Callegaro, S.~Singh, M.~Levorato, and F.~Restuccia, ``Bottlefit: Learning compressed representations in deep neural networks for effective and efficient split computing,'' in \emph{2022 IEEE 23rd International Symposium on a World of Wireless, Mobile and Multimedia Networks (WoWMoM)}.\hskip 1em plus 0.5em minus 0.4em\relax IEEE, 2022, pp. 337--346.

\bibitem{10.1145/3093337.3037698}
\BIBentryALTinterwordspacing
Y.~Kang, J.~Hauswald, C.~Gao, A.~Rovinski, T.~Mudge, J.~Mars, and L.~Tang, ``Neurosurgeon: Collaborative intelligence between the cloud and mobile edge,'' \emph{SIGARCH Comput. Archit. News}, vol.~45, no.~1, p. 615–629, apr 2017. [Online]. Available: \url{https://doi.org/10.1145/3093337.3037698}
\BIBentrySTDinterwordspacing

\bibitem{10.1145/3527155}
\BIBentryALTinterwordspacing
Y.~Matsubara, M.~Levorato, and F.~Restuccia, ``Split computing and early exiting for deep learning applications: Survey and research challenges,'' \emph{ACM Comput. Surv.}, vol.~55, no.~5, dec 2022. [Online]. Available: \url{https://doi.org/10.1145/3527155}
\BIBentrySTDinterwordspacing

\bibitem{salmani2023reconciling}
M.~Salmani, S.~Ghafouri, A.~Sanaee, K.~Razavi, M.~M{\"u}hlh{\"a}user, J.~Doyle, P.~Jamshidi, and M.~Sharifi, ``Reconciling high accuracy, cost-efficiency, and low latency of inference serving systems,'' in \emph{Proceedings of the 3rd Workshop on Machine Learning and Systems}, 2023, pp. 78--86.

\bibitem{ghafouri2024solution}
S.~Ghafouri, K.~Razavi, M.~Salmani, A.~Sanaee, T.~L. Botran, L.~Wang, J.~Doyle, and P.~Jamshidi, ``[solution] ipa: Inference pipeline adaptation to achieve high accuracy and cost-efficiency,'' \emph{Journal of Systems Research}, vol.~4, no.~1, 2024.

\bibitem{zhang2023effect}
X.~Zhang, M.~Mounesan, and S.~Debroy, ``Effect-dnn: Energy-efficient edge framework for real-time dnn inference,'' in \emph{2023 IEEE 24th International Symposium on a World of Wireless, Mobile and Multimedia Networks (WoWMoM)}.\hskip 1em plus 0.5em minus 0.4em\relax IEEE, 2023, pp. 10--20.

\bibitem{icfec2024}
\BIBentryALTinterwordspacing
M.~Mounesan, M.~Lemus, H.~Yeddulapalli, P.~Calyam, and S.~Debroy, ``Reinforcement learning-driven data-intensive workflow scheduling for volunteer edge-cloud,'' 2024. [Online]. Available: \url{https://arxiv.org/abs/2407.01428}
\BIBentrySTDinterwordspacing

\bibitem{stolaroff2018energy}
J.~K. Stolaroff, C.~Samaras, E.~R. O’Neill, A.~Lubers, A.~S. Mitchell, and D.~Ceperley, ``Energy use and life cycle greenhouse gas emissions of drones for commercial package delivery,'' \emph{Nature communications}, vol.~9, no.~1, p. 409, 2018.

\bibitem{konda1999actor}
V.~Konda and J.~Tsitsiklis, ``Actor-critic algorithms,'' \emph{Advances in neural information processing systems}, vol.~12, 1999.

\bibitem{mnih2016asynchronous}
V.~Mnih, A.~P. Badia, M.~Mirza, A.~Graves, T.~Lillicrap, T.~Harley, D.~Silver, and K.~Kavukcuoglu, ``Asynchronous methods for deep reinforcement learning,'' in \emph{International conference on machine learning}.\hskip 1em plus 0.5em minus 0.4em\relax PMLR, 2016, pp. 1928--1937.

\end{thebibliography}

\end{document}